\documentstyle[prb,aps,multicol,epsf]{revtex}
\begin{document}
\draft
\title{Doping dependence of the  exchange energies
 in bilayer
manganites: Role of orbital degrees of freedom}
 \author{G. Jackeli$^{1}$\cite{byline}, N.B. Perkins$^{2,3}$} 
\address{$^{1}$Centre de Recherches sur les Tr\`es Basses
Temp\'eratures, Laboratoire Associ\'e \'a l'Universit\'e Joseph Fourier,
\\ Centre National de la Recherche Scientifique, BP 166, 38042, Grenoble
C\'edex 9, France \\
$^{2}$Joint Institute for Nuclear Research, Dubna, Moscow region,
141980, Russia \\
$^{3}$ Max-Planck-Institut f$\ddot{u}$r
Physik komplexer Systeme, N$\ddot{o}$thnitzer Str. 38 01187 Dresden, Germany.}
\maketitle
\begin{abstract}
Recently, an intriguing  doping dependence of the exchange energies
in  the  bilayer manganites
${\rm La}_{2-2x}{\rm Sr}_{1 + 2x}{\rm Mn}_2{\rm O}_7$ 
has been observed in the  neutron scattering experiments.
The intra-layer exchange   only weakly changed with  doping 
while the inter-layer one drastically decreased. 
Here we propose a theory which accounts for these
experimental findings.  
We argue, that the observed striking doping dependence of the 
exchange energies can be attributed to the evaluation 
of the orbital level splitting with doping.
The latter is handled by the  interplay between 
Jahn-Teller effect  (supporting an axial orbital) and 
the orbital anisotropy of the electronic band in the bilayer structure
(promoting an in-plane orbital), 
which is  monitored by the Coulomb repulsion.
The presented theory, while being a mean-field type,  
describes well the experimental data and 
also gives the estimates of the several interesting 
 energy scales involved in the problem.
\end{abstract} 
\begin{multicols}{2} 
\narrowtext

The three-dimensional  (3D) cubic  manganites
${\rm R}_{1-x}{\rm B}_x{\rm MnO}_3$ (where 
${\rm R}$ is trivalent rare--earth and ${\rm B}$ is divalent alkaline ion, 
respectively) as well as the two-dimensional (2D) bilayer ones   
${\rm La}_{2-2x}{\rm Sr}_{1 + 2x}{\rm Mn}_2{\rm O}_7$ 
have attracted recent interest  not only due to the discovery  
of colossal magnetoresistance 
(CMR) in these compounds but also because of their rich and rather 
unusual physical properties.\cite{rev} 
At different doping concentration the manganese oxides  
exhibit a wide diversity of the ground states, and  
the small changes of  some external parameters of the system such as 
doping, chemical pressure, temperature or magnetic field can result in a 
drastic modification of the physical properties as well as cause  
the transition from one to another ground state. 
 
In fact, in very recent spin-wave measurements  on the  
bilayer manganites  ${\rm La}_{2-2x}{\rm Sr}_{1 + 2x}{\rm Mn}_2{\rm O}_7$ 
(in the doping range $x=0.3-0.5$) 
strongly anisotropic doping dependence of the 
inter-layer (within the bilayer) ($J_\perp$)  
and  of  the intra-layer ($J_\parallel$)  exchange couplings  
has been revealed.\cite{hirota,perring}   
At $x=0.3$ both exchange constants are  ferromagnetic (FM) and
of the same 
order $J_\perp$=5 meV, and $J_\parallel$=4 meV.\cite{hirota} 
With further doping  ($x>0.3$), 
 $J_\parallel$ changes very slightly while 
inter-layer exchange  rapidly decreases, 
by a factor  of four at $x=0.4$, 
and changes  sign at $x\sim 0.45$.

Within the double-exchange  (DE) picture FM  coupling  
between the nearest-neighbour (NN) $t_{2g}$ core spins 
is mediated by the hopping of $e_{g}$-electrons  
($d_{x^{2}-y^{2}}$ and $d_{3z^{2}-r^{2}}$  orbital states) and  
 scales as a kinetic energy of these electrons on a given bond. 
Since only $d_{3z^{2}-r^{2}}$ (axial) orbital   has a finite inter-layer 
transfer amplitude ($z$-axes is set perpendicular to a bilayer), 
the inter-layer 
exchange is {\it only} meditated by this orbital. 
While the intra-layer one is {\it mainly} 
determined by the $d_{x^{2}-y^{2}}$ (in-plane) 
orbital state, that has a highest 
intra-plane transfer amplitude.   
Therefore, above discussed decrease of the inter-layer exchange constant 
with doping can be ascribed to the change of the nature of occupied orbital state from mainly $d_{3z^{2}-r^{2}}$ character, 
at low doping $x<0.3$, to mostly  $d_{x^{2}-y^{2}}$ one at higher doping.
\cite{hirota,perring}
This qualitative picture is also consistent with the recent   
  x-ray magnetic Compton-profile 
measurements.\cite{koizumi}  The results of this experiment have shown
that upon doping (for  $0.35<x<0.45$ ) 
the electrons are mainly withdrawn from $d_{3z^{2}-r^{2}}$ orbital state 
and the occupancy of $d_{x^2-y^2}$ state remains practically the same. 
However, we point out, that the observed doping dependence of the
exchange constants and of the orbital occupancy, being interrelated, 
can not be understood within a rigid band picture
that assumes a constant value of orbital level splitting.
Interpretation  of the experimental findings  within
this picture would imply 
that only $d_{3z^2-r^2}$ band crosses the 
Fermi level and  is emptied out upon doping,
while the $d_{x^2-y^2}$ band lies below and shows a constant filling.
From the other hand,
if $d_{x^2-y^2}$ band does not cross Fermi surface, 
then it is no longer active to generate the intra-layer
ferromagnetic DE interaction leading to  $J_{\parallel}< J_{\perp}$, 
in contrast to experimental findings $J_{\parallel}> J_{\perp}$.

The aim of the present work is to  propose a mechanism that can  explain 
the 
experimental findings.
The paper is organized as follows: First we briefly discuss 
the structural aspects of the compound 
relevant for our study. Next we give  the basic physical ideas and
set up the minimal model Hamiltonian to provide a microscopic description.
We then present our results on the doping evaluation of the orbital 
occupancies and the exchange energies, and  
finally, compare the theoretical findings with
that of the experiments.
  
The compound  ${\rm La}_{2-2x}{\rm Sr}_{1 + 2x}{\rm Mn}_2{\rm O}_7$
 consists of  
bilayer slices of MnO$_6$ octahedra, separated 
 by insulating (La,Sr)$_2$O$_2$ layers that serve to decouple the bilayers 
 both electronically and magnetically. Therefore,  
as a first approximation the system can  be treated as composed of independent 
bilayers.  
One electron placed in the doubly degenerate $e_{g}$ level, makes the  
Mn$^{+3}$O$^{-2}_6$ complex Jahn-Teller (JT) active. 
The  experimental analysis of the crystal strutter  
has revealed, that JT distortion  results in  the   
elongation of ${\rm Mn}{\rm O}_{6}$ octahedra along $z$-axis. 
\cite{kubota} 
Five bonds (four equatorial and one apical with  
shared oxygen ion) are essentially identical. The elongation of octahedra  
occurs mainly because the apical bond with unshared oxygen atom is longer than the  other five bonds.  These type of JT distortion promotes the 
axially directed $d_{3z^2-r^2}$ orbital by lowering its on-site energy. 
The  dimensionless parameter describing  
 the static Jahn-Teller  
distortion can be defined as 
$\delta_{\text{JT}}\equiv (\langle d_{\text{Mn-O}}^{\text{apic}}\rangle-\langle d_{\text{Mn-O}}^{\text{equat}}\rangle)/\langle d_{\text{Mn-O}}^{\text{equat}}\rangle$, 
where $\langle d_{\text{Mn-O}}^{\text{apic(equat)}}\rangle$ is an
 averaged apical (equatorial) Mn-O length. 
The relative distortion $\delta_{\text{JT}}$ also gives the scale 
of JT induced  orbital level splitting. 
 
The physical picture we shall be based on is as follows:
Both the orbital occupancy and exchange energies strongly depend on
the orbital level splitting.
The latter consist of the two parts: JT induced gap and
correlation induced one.
The JT distortion is  largest at low doping and monotonically relaxes
upon doping
($\delta_{\text{JT}}\sim 0.036,~ 0.020,~\rm{and}~0.0$ for
$x=0.3,~0.4,~\rm{and}~0.5$, respectively).\cite{kubota}  
Therefore, at low doping the axial orbital, promoted by the JT distortion
is predominantly occupied. The matrix element of the 
 inter-orbital Coulomb energy does not depend on the character of predominantly occupied orbital state.  
Therefore,  on-site repulsion 
further supports already preferred orbital state 
and by enhancing  the orbital polarization reduces 
the Coulomb energy.\cite{coul} 
At low doping, the physics is mainly determined locally and 
kinetic energy term is less important.
However, upon increasing the carriers number the JT induced gap decreases
while  the kinetic energy starts to play more dominant role.
In the bilayer system, the latter promotes the in-plane orbital 
state, that  has the highest intra-layer transfer amplitude 
and by forming the wider band can lower the kinetic energy of the system. 
Therefore, one would expect that at some doping $x=x_{c}$ 
the possible energy gain due to  the kinetic energy will 
overcome the crystal-field splitting of the orbital  
levels.  Therefore, for $x>x_{c}$, it  becomes energetically more 
favorable to lower the energy of the  in-plane orbital and hence, 
 correlation induced splitting will  change the sign. 
As will be shown below, the doping dependence of the
orbital level splitting, resulted from the discussed interplay,
gives the consistent description  of the experimental findings.

The minimal model Hamiltonian which describes above discussed 
scenario consists of the two contributions $H=H_{\text{el}}+H_{\text{sp}}$, 
where $H_{\text{el}}$ and $H_{\text{sp}}$
describe charge and spin degrees of freedom, respectively.\cite{scint}
We retain only the preferred spin component of the fermionic operators,
 in the fully polarized (semi-metallic) FM case and also assume that
all the other degrees of freedom are integrated out to give effective model
parameters relevant to the low energy physics. 
The electronic part of the Hamiltonian can be written as:

\begin{eqnarray}
&H_{\text{el}}&=
-
\!\!\!\sum_{ij,\alpha ,\beta}
t_{{\parallel},ij}^{\alpha\beta}\left[ 
d_{i\alpha}^{\dagger}d_{j\beta}+
{\bar d}_{i\alpha}^{\dagger}{\bar d}_{j\beta}\right]
-\!\!\!\sum_{i,\alpha ,\beta} 
t_{{\perp}}^{\alpha\beta}
\left[d_{i \alpha  }^{\dagger}{\bar d}_{i\beta}+\text{H.c.}
\right]
\nonumber\\
&+&
\!\!\sum_{i,\alpha}
\left[(-1)^{\alpha}\Delta_{\text{JT}}-\mu\right]
\left[n_{i\alpha}+{\bar n}_{i\alpha}\right]
+\!\!\sum_{i}
U_{\text{eff}}\left[n_{i1}n_{i2}+
{\bar n}_{i1}{\bar n}_{i2}\right]\nonumber\\
\label{1}
\end{eqnarray}
The first and the second term  of Eq.(\ref{1}) describe an 
intra- and inter-layer electron hopping between the two $e_{g}$ orbitals of the  
NN Mn-ions, respectively. Index $i$ numbers the unit cell composed by
two Mn sites,  $d_{i \alpha  }$ and ${\bar d}_{i\alpha}$  are the electron
annihilation operators on these two sites. The orbitals $d_{3z^{2}-r^{2}}$ and $d_{x^{2}-y^{2}}$ correspond to
$\alpha(\beta)=1$ and 2, respectively. Intra- and inter-layer transfer 
matrix elements  are given by
\begin{eqnarray}
t_{{\parallel},x(y)}^{\alpha \beta}=
t\left(\!\!\begin{array}{cc}
1/4 &\!
\mp\sqrt{3}/4\\
\mp\sqrt{3}/4&\!3/4
\end{array}
\!\!\right)~,\;\;\;
t_{{\perp}}^{\alpha \beta}=
t\left(\!\!\begin{array}{cc}
1 &\!\;\;
0\\
0&\!\;\;0
\end{array}
\!\!\right)\;\;\;  .
\label{2}
\end{eqnarray}
The electron density operator on a given orbital state is denoted by $n_{i\alpha}$,
$\mu$ is chemical potential, $\Delta_{\text{JT}}=g\delta_{\text{JT}}$  is JT induced orbital level splitting, $g$  is properly normalized coupling constant  
of electrons with JT active phonon modes, and
$\delta_{\text{JT}}$ is a measure of JT distortion discussed above.
In the present paper we do not attempt to calculate JT distortion by
minimizing the total energy of the system (including lattice elastic energy)
and model the dimensionless parameter $\delta_{\text{JT}}$ by the following
doping dependence $\delta_{\text{JT}}=0.17(0.5-x)$, that reasonably reproduces
the experimental data.\cite{kubota}

The last term in Eq.(\ref{1}) represents an effective, 
relevant for the low energy physics, on-site inter-orbital Coulomb 
repulsion  between  electrons. 
We assume that $U_{\text{eff}}$  has been already renormalized  
due to the many-body effects
and properly screened in the metallic state. Therefore, one expects that
$U_{\text{eff}}$ is much smaller than the bear ionic value of Hubbard $U$.
\cite{U}
We treat this term within the  mean-filed (MF) 
approximation by introducing the MF parameter describing the anisotropy
 of the orbital occupancy
$\delta n=\langle n_{1}\rangle-\langle n_{2}\rangle$. 
This introduces additional splitting of local orbital levels due to  
the electron-electron interaction. Therefore  the total splitting
of orbital levels will be determined by  both JT effect and  electron-electron
repulsion 
 with following value: $2\Delta=2\Delta_{\text{JT}}+U_{\text{eff}}\delta n$.
We  emphasize, that the nonzero value of MF parameter
$\delta n$ does not imply any symmetry breaking long range orbital order
and the symmetry of Hamiltonian remains tetragonal.
Even in the non-interacting case and absence of any JT distortion one finds
$\delta n \not =0 $. The band structure with tetragonal symmetry
brakes the local cubic symmetry and  promotes planar $d_{x^2-y^2}$ orbital
with wider band leading to $\delta n < 0 $. 
In the orbital pseudo-spin language the ground state
can be represented by nonzero value of the axial component of the orbital 
pseudo-spin 
$\langle\tau_i^z\rangle=\langle (n_{i1}-n_{i2})/2\rangle\not =0$  due to the nonzero
 pseudo-magnetic field produced by the 2D structure of the system, and  with no order
in the pseudo-spin basal plane $\langle\tau_i^{x(y)}\rangle=0$.

The resulted MF Hamiltonian is bilinear and can be diagonalized by two 
subsequent canonical transformations.\cite{1} 
First step is to  introduce the bonding and anti-bonding states
as 
$a(b)_{\alpha(\beta){\bf k}}=[
d_{\alpha(\beta){\bf k} }\mp{\bar d}_{\alpha(\beta){\bf k}}  ]/\sqrt{2}$ 
and  decouple
the MF Hamiltonian in two parts each consisting of the  two orbitals. 
The next  transformation   
$A(B)_{1{\bf k}}=
u^{a(b)}_{{\bf k}}a(b)_{1{\bf k}}+v^{a(b)}_{{\bf k}}a(b)_{2{\bf k}}
$ diagonalizes the two-band Hamiltonian bringing it into the form:
$H=\sum _{{\bf k}\alpha} 
\left[\varepsilon^a_{\alpha{\bf k}}
A_{\alpha{\bf k}}^{\dagger}A_{\alpha{\bf k}}+
\varepsilon^b_{\alpha{\bf k}}
B_{\alpha{\bf k}}^{\dagger}B_{\alpha{\bf k}}
\right].$
The coherence factors and eigen-frequencies are given by
\begin{eqnarray}
         u[v]^{a(b)}_{{\bf k}}&=&[\text{sgn}(\varepsilon^{12}_{{\bf k}})]
\frac{1}{\sqrt{2}}
\left[1+[-]\frac{\varepsilon^{11}_{a(b){\bf k}}-
\varepsilon^{22}_{{\bf k}}}{\varepsilon^{a(b)}_{1{\bf k}}-
\varepsilon^{a(b)}_{2{\bf k}}}\right]^{\frac{1}{2}},\nonumber\\
\varepsilon^{a(b)}_{1(2){\bf k}}&=&
\frac{\varepsilon^{11}_{a(b){\bf k}}+\varepsilon^{22}_{{\bf k}}}{2}
\pm
\sqrt{\frac{\left [\varepsilon^{11}_{a(b)\bf k}-\varepsilon^{22}_{\bf k}\right ]^2}{4}+
\left [~\varepsilon^{12}_{{\bf k}}~\right ]^2 }
\label{6}
\end{eqnarray}
with the following notations
$\epsilon_{a(b){\bf k}}^{11}=
-t\gamma_{\bf k}
\pm t-\mu +\Delta$, $\epsilon_{{\bf k}}^{22}=
-3t\gamma_{\bf k}-\mu-\Delta$,
$\epsilon_{{\bf k}}^{12}=
-\sqrt{3}t\bar{\gamma}_{\bf k}$, and 
$\gamma(\bar{\gamma})_{\bf k}=[\cos k_{x}\pm\cos k_{y}]/2$.

The above introduced MF parameter $\delta n$ and chemical potential 
$\mu$ for a given doping 
can be obtained from the system of  self-consistent equations which in terms of the
obtained  eigen-states of MF Hamiltonian reads as:
\begin{eqnarray}
 n&=&\frac{1}{2N}\sum_{{\bf k}}\left\{  
n(\varepsilon^a_{1{\bf k}})+n(\varepsilon^a_{2{\bf k}})+
[b\rightarrow a]\right\}
\label{8}\\
\delta n&=&\frac{1}{2N}\sum_{\bf k} \left\{ 
[(1-
2(v^{a}_{{\bf k}})^2 ]
[n(\varepsilon^a_{1{\bf k}})-n(\varepsilon^a_{2{\bf k}})]
+[b\rightarrow a]\right\}
\nonumber
\end{eqnarray}
where $n(\varepsilon)$ is the Fermi distribution function.

Lets us now discuss the spin degrees of freedom of the system.
In the DE exchange limit (Hund's coupling $J_{\rm H}\gg W$ carriers band-width)
the spin  subsystem in the two orbital model 
can be also  mapped to an effective  NN Heisenberg model 
$H=-\sum_{i,\delta_{\parallel}}J_{\parallel}{\bf S}_i
{\bf S}_{i+\delta_{\parallel}}-
\sum_{i,\delta_{\perp}}J_{\perp}{\bf S}_i
{\bf S}_{i+\delta_{\perp}}$
with  effective intra-layer $J_{\parallel}$ and inter-layer 
$J_{\perp}$ FM exchange couplings defined as 
$J_{\parallel}=J_{\parallel}^{DE}-J$, $J_{\perp}=J_{\perp}^{DE}-J$, and
 $J$ is the antiferromagnetic super-exchange constant between the $t_{2g}$ spins. Ferromagnetic intra- and inter-layer DE energies are given by
\begin{eqnarray}
J_{\parallel}^{DE}\!\!=\sum_{\alpha ,\beta}
t_{\parallel}^{\alpha\beta}
\frac{\langle d_{i,\alpha}^{\dagger}d_{j,\beta}\rangle}{2S^{2}},\;\;\;
J_{\perp}^{DE}\!\!=\sum_{\alpha ,\beta}
t_{\perp}^{\alpha\beta}
\frac{\langle d_{i,\alpha}^{\dagger}\bar{d}_{j,\beta}\rangle}{2S^{2}}.
\label{11}
\end{eqnarray}  

With the above formulated scheme we proceed as follows.
First, we solve the MF equations (\ref{8}) to determine, the orbital level splitting $\Delta$ and chemical potential $\mu$ for a given doping.
Then, expressing  the Eq.~\ref{11} for the   exchange constants
in terms of the eigen-states of the MF Hamiltonian we end up with
the intra- and inter-layer exchange energies for a given doping.
Model parameters are fixed in a way to reproduce the experimental results.

\begin{figure}
\epsfysize=55mm
\centerline{\epsffile{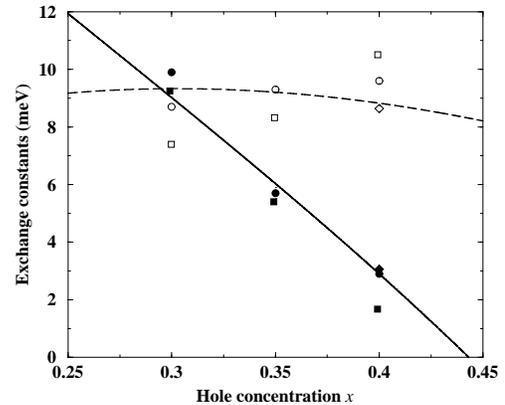}}
\caption{
Inter- ($SJ_{\perp}$) and intra-layer ($SJ_{\parallel})$
exchange constants  as a function of hole concentration
(solid and dashed line, respectively).
The filled (open) circles, squares, and diamonds are 
inter-(intra-)layer exchange constants from neutron scattering data
of Refs. 2, 3, and 9, respectively.}
\label{f1}
\end{figure}

In Fig.~\ref{f1} the calculated
doping dependence of intra- and inter-layer exchange couplings
is presented together with the experimental data from
 Refs.\onlinecite{hirota,perring,tapan} (see also Ref.\onlinecite{jpar}).
The best fit to the data has been achieved for  $t=0.18~\rm{eV}$, 
$U_{\text{eff}}=0.7~\rm{eV}$, $g=0.5~\rm{eV}$,
and $2SJ=11~\rm{meV}$ (see also discussion below).
The hopping integral sets the overall energy scale. The value 
of  electron-phonon coupling strength fixes  the doping  at which
$J_{\parallel}=J_{\perp}$ and system shows the isotropic behavior.
The interaction term  $U_{\text{eff}}$ determines the steepness of the drop of inter-plane exchange 
constant and  the AFM superexchange $J$ shifts  rigidly the whole picture relative to the $Y$ axis of Fig. 1.
Therefore,  despite the number of independent parameters all of them can be unambiguously extracted from the fitting.

As it is seen from Fig. \ref{1} the intra-layer exchange is practically unaffected 
by doping while the inter-layer one is dramatically reduced.
When  inequality $J_{\perp}^{DE}<J$ becomes valid, the 
antiferromagnetic super exchange prevail ferromagnetic DE signaling the instability of the FM ground state of the bilayer.
In fact, neutron diffraction study on $x=0.5$ sample revealed the
A-type AFM ordering.\cite{kubota}

\begin{figure}
      \epsfysize=55mm
      \centerline{\epsffile{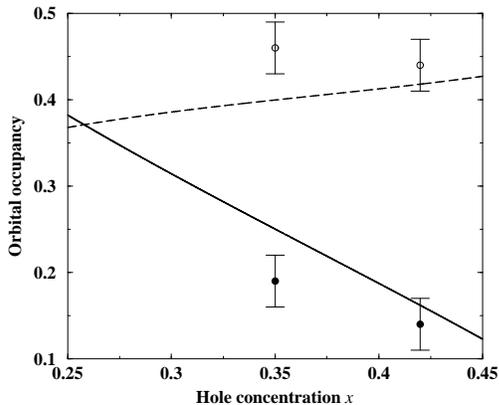}}
\caption{Occupancy of $d_{3z^2-r^2}$ (solid line) and $d_{x^2-y^2}$ (dashed
line) orbitals as a function of hole concentration.
The open (filled) circles are the experimental data for 
$d_{x^2-y^2}$ ($d_{3z^2-r^2}$) orbital occupancy from Ref. 4.
} \label{f2}
\end{figure}

As it follows, the above discussed doping dependence of the exchange constants
is in one to one correspondence with the doping dependence of the 
orbital occupancy. The latter is shown in Fig. 2 calculated for the same values
of the model parameters. The  experimental data from Ref.\onlinecite{koizumi}
is also presented on the same plot.
The  in-plane   
 $d_{x^{2}-y^{2}}$ orbital (dashed line in Fig. \ref{f2}) 
is predominantly occupied and shows weak doping dependence
in almost whole presented range of the hole concentration.
While axially directed $d_{3z^{2}-r^{2}}$  orbital 
(solid line in  Fig. \ref{f2}) is emptied-out linearly with doping, i.e.
all the doped hole in this doping range resides on  $d_{3z^{2}-r^{2}}$ orbital.

\begin{figure}
      \epsfysize=65mm
      \centerline{\epsffile{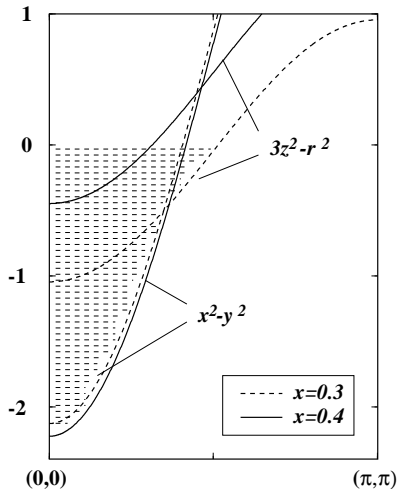}}
\caption{Bonding ${3z^2-r^2}$  and ${x^2-y^2}$ 
bands (in units $t=1$) referred to chemical potential for hole doping $x=0.3$ (dashed
line)  and $x=0.4$ (solid line).} 
\label{f3}
\end{figure}

The above discussed evaluation of the orbital occupancy upon doping 
 can be understood as follows.
In the  doping range $x<x_{c}$, ($x_{c}\simeq 0.26 $ being the 
hole concentration at which both orbitals are equally occupied [see Fig.~\ref{f2}])
the JT distortion is large and stabilizes the axially directed orbital.
With  doping JT induced gap weakens and  at $x=x_{c}$
the kinetic energy term supporting the in-plane orbital state takes over.
Therefore, $\delta n$ changes sign  at  $x=x_{c}$
 and  the $d_{x^{2}-y^{2}}$ orbital starts 
to be predominantly populated. For   $x>x_{c}$  
correlation induced splitting $2\Delta_{\rm cor}=U\delta n$ is negative
and   supports in-plane orbital.
With farther doping, the orbital anisotropy 
$\delta n$ increases, leading to increase of correlation induced splitting.
 Therefore, the axial band is pushed up relative to the in-plane band 
upon doping, while the   chemical potential is practically  pinned.
Therefore, all the doped holes go to the upper $d_{3z^{2}-r^{2}}$ band
and its occupancy linearly decrease upon doping, while the population
of lower $d_{x^2-y^2}$ band is only weakly changed.

This doping dependence of the band structure is explicitly shown  
in Fig.~\ref{f3}, where the bonding $d_{x^2-y^2}$ and $d_{3z^2-r^2}$ bands 
are presented for hole concentration $x=0.3$ (dashed line) and $x=0.4$ (solid line).

Let us know discuss the model parameters used to 
reproduce the experimental data.
The value of the  hopping integral $t=0.18~\rm{eV}$ is in the range 
estimated for manganese oxides $t\simeq 0.1 \sim 0.3~\rm{eV}$.\cite{kugel}
We also point out, that some of the authors estimated hopping integral to be
higher $t\sim 0.7\rm{eV}$.\cite{t}  
Moderate value of the effective inter-orbital Coulomb repulsion
$U_{\text{eff}}=0.7~\rm{eV}$ is sufficient to give reasonable fit to the data, 
which justifies the  MF treatment adopted here.\cite{U} 
The AFM superexchange energy of $t_{2g}$ spins remarkably coincides with 
one ($2SJ=10\rm{meV}$) estimated from the spin-wave data of  
${\rm Nd}_{0.45}{\rm Sr}_{0.55}{\rm Mn}{\rm O}_{3}$.\cite{1}
Another important parameter is the JT splitting of the orbital states,
that is difficult to directly detected experimentally.
For $x=0.3$, with the above estimate of the electron-phonon coupling constant
($g=0.5$ eV)\cite{g} 
one obtained the JT splitting $2\Delta_{JT}\simeq 0.035~\rm{eV}$.
Hence, in the doping regime considered here, the JT binding energy is much smaller than the carriers band-width. This explains why the polaronic effects,
not considered in the present paper, can be ignored.

The authors would like to thank  
T. Chatterji, N. M.  Plakida, V. Yu. Yushankhai  and 
M. E. Zhitomirsky  for useful  discussions.
G. J. acknowledges support from a  
{\it Bourse Post-Doctoral du  minist\`ere de l'Education nationale, 
                     de la recherche et de la technologie}
and N. B. P. acknowledges support from the 
 Visitor Program of MPI-PKS, Dresden.
The partial support by  INTAS program, Grant No. 97-11066,
 is also acknowledged. 


\end{multicols}

\begin{references}
\bibitem[*]{byline} On leave from E. Andronikashvili Institute of Physics,
Georgian Academy of Sciences, Tbilisi, Georgia. 
\bibitem{rev} 
For a review, see, for example,
E. Dagotto, T. Hotta and A. Moreo,
Phys. Rep. {\bf 344}, 1 (2001); 
E.L. Nagaev, {\it ibid}. {\bf 346}, 387 (2001). 
\bibitem{hirota} 
K. Hirota, S. Ishihara, H. Fujioka, M. Kubota, H. Yoshizawa,
Y. Morimoto, Y. Endoh, and S. Maekawa,
Phys. Rev. B {\bf 65}, 064414 (2002).
\bibitem{perring} T.G.Perring, D.T. Adroja, 
G. Chaboussant, G. Aeppli, T. Kimura, and Y. Tokura, 
Phys. Rev. Lett. {\bf 87}, 217201 (2001).
\bibitem{koizumi}
A. Koizumi, S.Miyaki, Y. Kakutani, H. Koizumi, N. Hiraoka, K. Makoshi, and
N. Sakai, K. Hirota, and Y. Murakami,
Phys. Rev. Lett. {\bf 86}, 5589 (2001). 
\bibitem{kubota}
M. Kubota, H. Fujioka, K. Hirota, K. Ohoyama, Y. Morimoto, H. Yoshizawa, and
Y. Endoh,
 J. Phys. Soc. Jpn. {\bf 69}, 1606 (2000).
\bibitem{coul} In the fully  polarized ferromagnetic phase
  the role of 
intra-orbital Coulomb  
repulsion is insignificant and can be dropped out. 
\bibitem{scint} Since, we only consider the  the effect of orbital degree's of freedom on the exchange energies, 
 the  interaction term between the spin and charge degrees of freedom is dropped out from the model Hamiltonian. 
\bibitem{U}
For a detailed  discussion on this issue see K. H. Ahn and A. J. Millis, 
Phys. Rev. {\bf 61}, 13 545 (2000); I. Solovyev, N. Hamada, and K. Terakura,
{\it ibid}. {\bf 53}, 7158 (1996).
\bibitem{1} G. Jackeli, N. B. Perkins and N. M. Plakida, Phys. Rev. B {\bf 62}, 372 (2000); {\bf 64}, 092403 (2001).
\bibitem{tapan} T. Chatterji, L. P. Regnault, P.Thalmeier, 
R. Suryanarayanan, G. Dhalenne, and A. Revcolevschi,
Phys. Rev. B {\bf 60}, R6965 (1999).
\bibitem{jpar} The slight discrepancy between the values of the 
intra-layer exchange reported by the various experimental groups 
is probably due  to the fact, that the different momentum regions of the spin--wave spectra 
have been used by those groups to extract  
$J_{\parallel}$. The inter-layer exchange can be more accurately determined, since it sets the momentum independent splitting of bonding and anti-bonding spin-wave spectra.
\bibitem{kugel} K. I. Kugel and D. I. Khomskii, Sov. Phys. Usp.
{\bf 25}, 231 (1982). Slightly different estimate $t=0.2-0.5$(eV) has been
reported in Ref.1. 
\bibitem{t} See, for example, Y.-R. Chen and P. B. Allen, Phys. Rev. B {\bf 64}, 064401 (2001).
\bibitem{g} The strength of electron-phonon coupling estimated here is close
to the lower boundary of the estimation given by 
A. J. Millis, Phys. Rev. B {\bf 53}, 8434 (1996) and agrees with that
reported by M. O. Dzero, L. P. Gor'kov,  and V. Z. Kresin,
Solid State Comm., {\bf 112}, 707 (1999).
\end{references}
\end{document}